**Title: Reduced structural connectivity between left auditory thalamus and the motion-sensitive planum temporale in developmental dyslexia**

**Short Title:**
**Reduced left MGB - mPT connectivity in dyslexia**

**Authors:**
Nadja Tschentscher[a,b]
Anja Ruisinger[a]
Helen Blank[c]
Begoña Díaz[d]
Katharina von Kriegstein[a,e]

[a]Max Planck Institute for Human Cognitive and Brain Sciences, Leipzig, Germany

[b]Research Unit Biological Psychology, Department of Psychology,
Ludwig-Maximilians-Universität Munich, Leopoldstr. 13, 80802 Munich, Germany

[c]University Medical Center Hamburg-Eppendorf, Center for Experimental Medicine
Institute of Systems Neuroscience, Martinistraße 52, 20246 Hamburg, Germany

[d]Center for Brain and Cognition, Departament de Tecnología de les Comunicacions,
Universitat Pompeu Fabra, Barcelona, Spain

[e]Department of Psychology, Technische Universität Dresden, Dresden, Germany

**Corresponding Author:**
Nadja Tschentscher
Max Planck Institute for Human Cognitive and Brain Sciences,
Stephanstraße 1A, 04103 Leipzig, Germany
E-mail: tschentscher@cbs.mpg.de
Phone: +49 151 681 09 854

Number of pages: 38
Word count abstract: 246
Word count introduction: 650
Word count discussion: 1369
Number figures main text: 8

Major classification: **Biological Sciences**
Minor classification: **Psychological and Cognitive Sciences**
Conflict of interest: The authors declare no competing financial interests.

# Abstract

Developmental dyslexia is characterized by the inability to acquire typical reading and writing skills. Dyslexia has been frequently linked to cerebral cortex alterations; however recent evidence also points towards sensory thalamus dysfunctions: dyslexics showed reduced responses in the left auditory thalamus (medial geniculate body, MGB) during speech processing in contrast to neurotypical readers. In addition, in the visual modality, dyslexics have reduced structural connectivity between the left visual thalamus (lateral geniculate nucleus, LGN) and V5/MT - a cerebral cortex region involved in visual movement processing. Higher LGN-V5/MT connectivity in dyslexics was associated with the faster rapid naming of letters and numbers (RANln), a measure that is highly correlated with reading proficiency. We here tested two hypotheses that were directly derived from these previous findings. First, we tested the hypothesis that dyslexics have reduced structural connectivity between the left MGB and the auditory motion-sensitive part of the left planum temporale (mPT). Second, we hypothesized that the amount of left mPT-MGB connectivity correlates with dyslexics RANln scores. Using diffusion tensor imaging based probabilistic tracking we show that male adults with developmental dyslexia have reduced structural connectivity between the left MGB and the left mPT – confirming the first hypothesis. Stronger left mPT-MGB connectivity was not associated with faster RANnl scores in dyslexics, but in neurotypical readers. Our findings provide first evidence that reduced cortico-thalamic connectivity in the auditory modality is a feature of developmental dyslexia, and that it may also impact on reading related cognitive abilities in neurotypical readers.

## Significance Statement

Developmental dyslexia is one of the most widespread learning disabilities. While previous neuroimaging research mainly focused on pathomechanisms of dyslexia at the cerebral cortex level, several lines of evidence suggest an atypical functioning of subcortical sensory structures. By means of diffusion tensor imaging, we here show that dyslexic male adults have reduced white matter connectivity in a cortico-thalamic auditory pathway between the left auditory motion-sensitive planum temporale (mPT) and the left medial geniculate body (MGB). Connectivity strength of this pathway was associated with measures of reading fluency in neurotypical readers. This is novel evidence on the neurocognitive correlates of reading proficiency, highlighting the importance of cortico-subcortical interactions between regions involved in the processing of spectrotemporally complex sound.

## Introduction

Developmental dyslexia is characterized by the inability to develop neurotypical levels of reading and writing skills, despite unimpaired fluid intelligence and adequate educational opportunities (Peterson and Pennington, 2012, 2015). It is the most widespread learning disability with a prevalence of 5-10 percent in children, and often leads to severe emotional and social difficulties (Carroll and Iles, 2006).

Most neuroscientific research has investigated pathomechanisms of developmental dyslexia – which we will refer to as dyslexia in the following – at the cerebral cortex level (for review, see Vandermosten et al., 2012). However, there is also evidence of dyslexia related alterations in subcortical regions in the left auditory thalamus (medial geniculate body, MGB) (Díaz et al., 2012), and auditory brainstem (Chandrasekaran et al., 2009; Chandrasekaran and Kraus, 2010). Examinations of post-mortem brains showed histological changes in the left MGB of several cases with dyslexia (Galaburda et al., 1994; cf. Stein, 2001). It is a long-standing but never tested hypothesis that these changes are associated with altered fibre connectivity between the left MGB and those cerebral cortex areas that show alterations in dyslexia (cf. Galaburda et al., 1994).

We here tested the specific hypothesis that dyslexics have reduced structural connectivity between the left MGB and those parts of the left auditory association cortex (planum temporale, PT) (Westbury et al., 1999) that are involved in the processing of auditory motion. We will abbreviate this region mPT (motion-sensitive planum temporale). Our focus on the mPT rests on two strands of evidence. First, the PT has been implicated in dyslexia as part of a left-hemispheric dysfunctional language system (for review, see Shapleske et al., 1999; Altarelli et al., 2014). Structural MRI analyses showed atypical inter-hemispheric symmetry of PT volumes (Altarelli et al., 2014), and post-mortem brain analyses reported histological alterations in the PT of dyslexics (Galaburda et al., 1985; Humphreys et al., 1990). Critically, those histological alterations have not been reported for primary auditory (A1) cortices (for review, see Eckert, 2004). Second, in the visual modality, reduced left-hemispheric structural connectivity has been observed between the visual thalamus (lateral geniculate nucleus, LGN) and middle temporal area V5/MT, while connectivity between the LGN and the primary visual cortex (V1) was at a neurotypical level (Müller-Axt et al., 2017). V5/MT is an extrastriate visual cortex region implicated in visual motion processing (Britten et al., 1992; Zeki, 2015). The functionally most equivalent region to V5/MT in the auditory modality is the PT, which houses a key region for auditory motion processing (Warren et al., 2002; Alink et al., 2012), i.e. the region which we here call mPT. The PT is tonotopically organised (Langers, 2014), and animal tracing studies (Bajo et al., 1995; Winer et al., 2001; Lee and Winer, 2008) reported direct fibre connectivity between tonotopically organized regions (Langers and van Dijk, 2012; Langers, 2014) and the MGB.

Using diffusion tensor imaging based probabilistic tractography, we here compared male adults with dyslexia (N=12) with matched neurotypicals (N=12). First, we tested whether connectivity strength between the left mPT and the left MGB is weaker in dyslexics than in neurotypicals. Second, we expected that connectivity strength between the left mPT and the left MGB in dyslexics correlates with the rapid automatized naming of letters and numbers (RANln), and with reading comprehension. The RANln predicts reading abilities (Semrud-Clikeman et al., 2000; Miller et al., 2006; cf. Lervåg and Hulme, 2009; for review, see Norton and Wolf, 2012). In dyslexic adults, lower RANln and reading comprehension scores were associated with reduced fMRI responses in the left MGB (Díaz et al., 2012). The RANln also correlated with connectivity strength in the left LGN-V5/MT pathway (Müller-Axt et al., 2017). We tested the specificity of the expected left mPT-MGB connectivity reduction by assessing the right-hemispheric white matter connectivity between MGB and mPT, and MGB's connectivity with the primary auditory cortex (A1). We also explored MGB's connectivity with the immediately preceding nucleus in the auditory brainstem, the inferior colliculus (IC).

## Materials and Methods

### 1. Participants

Data from two groups of healthy male native German speakers without any history of neurological and psychiatric diseases were analyzed. The group of subjects with reading and writing impairments (dyslexia group) included 12 participants, of which 6 had been formally diagnosed with dyslexia, while the other 6 reported severe reading and spelling difficulties since childhood. The neurotypical group included 12 participants with average reading and spelling abilities, and was matched in age, sex, educational level, handedness, and non-verbal IQ (Raven, 1998) to the dyslexia group. For the same participant groups, a dyslexia specific reduction in left MGB responses during speech processing has been reported (Díaz et al., 2012), as well as a reduction in structural connectivity between the left LGN and V5/MT (Müller-Axt et al., 2017).

A formal behavioral assessment on reading speed and comprehension (Schneider et al., 2007), spelling (Kersting and Althoff, 2004), and skills of rapid automatized naming (i.e., RANln) (Denckla and Rudel, 1976) confirmed the group assignments: lower scores on spelling, reading speed, and comprehension were observed in the dyslexia group, as well as longer reaction times in RANln (Table 1). A median-split analysis, based on the average scores of spelling, reading speed, and reading comprehension revealed the same group assignments as defined a priori based on clinical diagnoses of dyslexia, and self-reports of participants on their reading and spelling abilities. The diagnostic test scores and social demographic variables are summarized in Table 1. Written informed consent was obtained from all participants before data acquisition. The study was approved by the ethics committee of the Medical Faculty, University of Leipzig, Germany.

[Please insert Table 1 here]

**2. Data acquisition of diffusion-weighted and anatomical T1-weighted images**

We acquired diffusion-weighted images (dMRI) on a 3 Tesla Magnetom Tim Trio MRI system with a 32-channel head coil (Siemens Healthineers, Erlangen, Germany), using a twice-refocused spin-echo echo-planar imaging (EPI) sequence (TE = 100ms, TR = 12.9 s, FOV = 220 x 220 mm², voxel size = 1.72 x 1.72 x 1.7 mm³). Eighty-eight axial slices were obtained, covering the whole brain without inter-slice gap. Diffusion weighting was isotropically distributed along 60 diffusion-encoding gradient directions with a b-value of 1000 s/mm². For off-line motion correction, seven interspersed anatomical reference images were acquired without diffusion-weighting (b-value = 0 s/mm²): one in the beginning, as well as one after each of the blocks of 10 diffusion-weighted images. Fat saturation was applied using a spectral saturation pulse. Generalized autocalibrating partially parallel acquisitions (GRAPPA; Griswold et al., 2002) was used with an acceleration factor of 2, as well as partial Fourier imaging of 6/8 to accelerate the dMRI acquisition. The dMRI sequence took approximately 16 min. A T1-weighted structural 3D image was acquired as anatomical reference on the same MRI system (MPRAGE, TE = 3.46 ms, TR = 1300 ms, TI = 650 ms, flip angle = 10°, 1mm isotropic resolution, two averages).

### 3. Preprocessing of dMRI data

FSL (FMRIB Software Library, University of Oxford http://www.fmrib.ox.ac.uk/fsl) was used to estimate motion correction parameters for the dMRI data based on the seven reference images without diffusion-weighting and rigid-body registration (Jenkinson et al., 2002). Motion correction parameters were interpolated for all 67 volumes and combined with a global registration to the T1 anatomy (in AC/PC space) using rigid-body registration. The estimated motion correction parameters were then used to correct the gradient directions of each dMRI volume. The registered dMRI volumes were sampled with an isotropic voxel resolution of 1.72 mm and the background was masked with the skull-stripped T1-image. A diffusion tensor was fitted to each voxel, and fractional anisotropy (FA) maps were computed (Basser et al., 1994; Basser and Pierpaoli, 1996).

### 4. Functional localizer for region of interest definition

To localize regions of interest (ROIs) in the auditory pathway, we used a functional localizer acquired on the same participant sample (cf. Díaz et al., 2012). Participants listened passively to blocks of auditory sentences that were separated by silence. The stimuli consisted of 40 five-word sentences that were semantically neutral (for example, "Der Junge trägt einen Koffer" meaning "The boy carries a suitcase") and syntactically similar (i.e., subject-verb-object), recorded from a 22-y-old male German speaker. The functional MRI (fMRI) data were acquired on a 3-T Magnetom TIM Trio (Siemens) by using a 32-channel head coil (Siemens Healthineers, Erlangen, Germany). A gradient-echo EPI sequence was applied (echo time = 30 ms, flip angle = 90°, acquisition bandwidth = 116 kHz, slice thickness = 2 mm, interslice gap = 1 mm, 42 slices, axial and ascending acquisition starting at the pontomedullary junction). A sparse sampling imaging protocol with cardiac gating was applied, which is thought to reduce the artifacts caused by the pulsatile motion of the brainstem (Hall et al., 1999; Thompson et al., 2006). The 82 brain volumes of each participant were analyzed with SPM8 (www.fil.ion.ucl.ac.uk/spm) and Matlab 7.10 (R2010a; MathWorks). Scans were realigned, unwarped, coregistered to the individual anatomical images, and normalized to Montreal Neurological Institute (MNI) standard stereotactic space. Images were spatially smoothed with a Gaussian smoothing kernel of 4 mm full width at half maximum for subcortical structures, and a smoothing kernel of 8 mm for cortical structures. A general linear model was applied using a boxcar function, convolved with a synthetic hemodynamic response function, to model the metabolic signal.

## 5. Definition of subcortical regions of interest

Subcortical ROIs were based on group-level peak coordinates from the functional localizer contrast "Sentences - Silence". Compared to the LGN (Müller-Axt et al., 2017), it is difficult to delineate the MGB on structural images (Devlin et al., 2006; Jiang et al., 2013; Tourdias et al., 2014). Devlin et al. (2006) have identified MGBs at the single subject level based on proton density (PD) images and a DTI approach, and Javad et al. (2014) defined MGBs on T1-weighted images; however there are so far no indications that these procedures yield more reliable information than a functional approach on the single participant level, or with random-effects group analyses with commonly used statistical thresholds (Jiang et al., 2013). Since localization at a single subject level was not possible in all participants in our study, we used a ROI based on the sample specific random effects coordinate obtained for the contrast "Sentences - Silence" (Díaz et al., 2012).

For the MGB, a spherical mask of 4 mm radius was defined around the statistical maxima present in the anatomical location of the left and right MGB (left [-15, -28, -5], and right [12, -28, -8] in MNI space). These coordinates have been previously used in fMRI analyses on the top-down modulation of the MGB in the same dyslexia sample (Díaz et al., 2012). The grey matter proportion of the mask matched the size of the reported grey matter volume of the MGB in previous in-vivo structural MRI as well as post-mortem analyses (Rademacher et al., 2002; Devlin et al., 2006; Javad et al., 2014), which reported mean left MGB volumes between 38 mm³ and 123 mm³ (see Figure 1 for masks and functional localizer clusters, and Table 2 for subcortical ROI volumes).

The MGB statistical maxima in the functional localizer were not exactly symmetric across hemispheres (left [-15, -28, -5], and right [12, -28, -8] in MNI space). Similar asymmetries in x and z direction have been previously observed by other neuroimaging studies (Devlin et al. (2006), left MGB [-14, -25, -6], right MGB [13, -25, -7]; Jiang et al. (2013), left MGB study 1 [-15, -25, -7], left MGB study 2 [-16, -26, -6], right MGB study 1 [13, -25, -6], right MGB study 2 [13, -26, -6]; Moerel et al. (2015), left MGB [-16, -23, -9], right MGB [14, -23, -9]).

[Please insert Figure 1 here]

Linear and non-linear registration was performed in FSL, using the default parameters, to bring the MGB masks into diffusion space. A reasonable placement of the masks on the single subject level was evaluated by visual inspection. To provide an example of this process, Figure 2 shows the placement of the MGB masks on the single subject T1 images of six randomly chosen subjects within the groups of dyslexics and neurotypicals. To extract grey and white matter, a binarized Fractional Anisotropy (FA) < 0.2 image of each participant was used in diffusion space (see Table 2 for grey and white matter ROI volumes).

[Please insert Figure 2 here]

For the IC, a spherical mask with a radius of 3 mm was created around peak coordinates from the functional localizer (left [-6, -34, -8], and right [6, -37, -5] in MNI space) (see Figure 3, yellow). The grey matter part of the mask matched the size of previously reported grey matter volumes of the IC (Sabanciogullari et al., 2013) (Table 2). Linear and non-linear registration was performed in FSL, using the default parameters, to bring the IC masks into diffusion space. Again, a binarized Fractional Anisotropy (FA) < 0.2 image of each participant was used to extract grey and white matter masks in diffusion space (see Table 2 for grey and white matter ROI volumes). Also for the IC, statistical maxima were not entirely symmetric (left [-6, -34, -8], and right [6, -37, -5] in MNI space). Asymmetries in IC coordinates have been observed in previous fMRI studies, however, whether this is a general feature of IC BOLD responses is unclear, as often IC coordinates are not reported (Melcher et al., 2000; De Martino et al., 2013; Ress and Chandrasekaran, 2013). There was also an alternative coordinate for the left IC at [-3, -37, -8] in the functional localizer contrast "Sentences - Silence". This coordinate had a slightly lower statistic (Z=4.25). We chose to report the results for [-6, -34, -8], due to its higher Z score (Z=4.35). However, we additionally performed analyses with ROIs based on the alternative coordinate [-3, -37, -8].

[Please insert Table 2 here]

### 6. Definition of cerebral cortex regions of interest

The cerebral cortex regions of interest in A1 and mPT were created in two steps. First, we created spherical masks around a functional MRI coordinate (for details see below). Second, these spherical masks were intersected with thresholded probabilistic volume based atlas masks to ensure that the ROIs did not exceed the anatomically defined regional boundaries of A1 and PT, respectively. For the A1 ROIs, the peak coordinates for the spherical masks were extracted from the functional localizer contrast "Sentences - Silence" in the left [-51, -16, 4] and right [42, -22, 7] hemispheres in MNI space (Figure 1). These coordinates have been used in previous fMRI analyses on the same subject sample (Díaz et al., 2012). To define the mPT, we used the coordinates reported by Alink et al. (2012), for the left [-53, -31, 12], and right [54, -29, 14] hemisphere in MNI space, respectively. We chose these coordinates for the intersections with volume-based atlas masks, because Alink et al. successfully decoded the direction of moving sounds from fMRI response patterns of bilateral PT volumes centered at these coordinates.

We created spherical masks (radius = 8 mm) around A1 and mPT peak coordinates. The A1 sphere was intersected with the thresholded probabilistic TE1.0 atlas mask from the Juelich Histological Atlas (Eickhoff et al., 2005). The mPT sphere was intersected with the PT mask from the Harvard-Oxford-Atlas (Desikan et al., 2006). We used these two different atlases, because the Juelich Atlas does not include the PT, and the Havard-Oxford-Atlas does not include a mask solely covering A1. Both atlases are implemented in FSL in MNI standard space with a voxel resolution of 1 mm$^3$. Mask volumes from probabilistic atlases were thresholded to the size of the approximate volumes of A1 (Rademacher et al., 2001; Artacho-Pérula et al., 2004) and the PT (Hirayasu et al., 2000; Ratnanather et al., 2013), as known from post-mortem studies and MRI segmentations. Intersections between these masks and the spherical volumes were calculated in FSL (see Figure 3 for masks, and Table 3 for ROI volumes). Linear and non-linear registration was performed in FSL, using the default parameters, to bring the masks into diffusion space. A binarized Fractional Anisotropy (FA) < 0.2 image of each subject was used to extract grey and white matter masks in diffusion space (see Table 3 for grey and white matter ROI volumes).

[Please insert Figure 3 and Table 3]

### 7. Probabilistic tractography

Anatomical connectivity was estimated using FDT (FMRIB's Diffusion Toolbox; (https://fsl.fmrib.ox.ac.uk/fsl/fslwiki/FDT). Voxel-wise estimates of the fiber orientation distribution were computed using BEDPOSTX (Bayesian Estimation of Diffusion Parameters

Obtained using Sampling Techniques) (Behrens et al., 2007). The distribution of up to two fiber orientations at each voxel was estimated based on the b-value and resolution of the dMRI data (Behrens et al., 2003). Probabilistic tractography was performed in native diffusion space using the PROBTRACKX2 module, to estimate the strength and the most likely location of a pathway between the respective seed and target areas. Probabilistic tracking from A1 and mPT volumes to the MGB, as well as between the MGB and the IC, was run for each hemisphere separately.

In cortico-subcortical tracking analyses, cortical regions (A1 and mPT) were set as seed masks. The left MGB was defined as waypoint and termination mask, so that only tracks were counted which reached the MGB, as well as terminated in the MGB (i.e., did not go further). We only computed the tractography from the cerebral cortex ROIs to the MGB, to ensure that possible non-dominant cortico-subcortical connectivity was detected as well, which might be missed by the algorithm when seeding in the MGB. Seeding in regions with relatively low anisotropy, such as the MGB, may lead to large uncertainty in fiber orientation (Jones, 2010).

Modified Euler streamlining was applied, while all other default parameters were kept. All analyses were done separately for each pair of seed and target region within each hemisphere. Tractography results were considered reliable when at least 10 of the generated sample streamlines reached the target, a threshold that has been used in previous probabilistic tracking studies (Heiervang et al., 2006; Makuuchi et al., 2009; cf. Blank et al., 2011; Müller-Axt et al., 2017). Less reliable connections were excluded from statistical analyses. Since the number of estimated streamlines for each pair of seed and target region is highly determined by the size of the seed, we corrected for the number of voxels in the respective seed mask. A connectivity index was calculated that reflects the connection strength between pairs of seed and target regions: the number of streamlines from a given seed that reach the target (*waytotals*) was log-transformed and divided by the log-transformed product of the generated sample streamlines in each seed voxel (5000) and the number of voxels in the respective seed mask ($V_{seed}$):

$$Connectivity \text{ Index} = \frac{log(waytotal)}{log(5000 * V_{seed})} \text{ ,}$$

The log-transformation increased the likelihood of gaining a normality distribution, which was tested before application of parametric statistics using the Shapiro-Wilk test (Royston, 1992).

For visualization purposes (in Figures 4, 6, and 7), the output images of the probabilistic tractography for each connection and subject were normalized by logarithmic transformation, and then divided by the logarithm (log) of the total number of generated streamlines in each seed mask. This was the same normalization procedure as described for computation of the connectivity index. The normalized tracts were transformed into MNI standard space, averaged within groups of dyslexics and neurotypicals, and thresholded to the same minimum value.

## 8. Statistical analyses

### 8.1 Group differences in connectivity strength

Statistical analyses were run in Matlab (Version R2016b). Individual participants' connections were considered as inconsistent if the connectivity index exceeded the threshold of 2.5 SDs above or below the group mean of their respective group. After exclusion of these outliers, connectivity indices were subjected to random-effects analyses. A two-sample t-test was used to test our first hypothesis that there is reduced connectivity in the left mPT-MGB connection in dyslexics in comparison to neurotypicals.

To test the specificity of the putative group differences in the left mPT-MGB connection, we performed two mixed-effects ANOVAs. First, we tested whether group differences in the mPT-MGB connection were present in the left mPT-MGB pathway, but not in the right mPT-MGB pathway. To do that, we performed a 2x2 mixed-effects ANOVA with the factors "hemisphere" (left versus right) and "group" (neurotypicals versus dyslexics). There were two reasons why we expected group differences in the left mPT-MGB pathway only: alterated BOLD responses of dyslexics in a speech processing task were restricted to the left MGB (Díaz et al., 2012), and post-mortem analyses revealed histological alterations in the left MGB while no such findings were reported for the right hemisphere (Galaburda et al., 1994; cf. Stein, 2001). In the second mixed-effects ANOVA with the factors "seed region" (left A1 versus left mPT) and "group" (neurotypicals versus dyslexics), we tested whether the putative group differences were present in the mPT-MGB connection, but not in the A1-MGB connection. The left A1-MGB connection was also tested for significant group differences, and putative hemispheric differences in the A1-MGB connection were assessed in an 2x2 mixed-effects ANOVA with the factors "hemisphere" (left versus right) and "group" (neurotypicals versus dyslexics). We did not expect any dyslexia specific abnormalities in the A1-MGB connection, since A1 as opposed to PT did not reveal histological alterations in post-mortem studies (for review, see Eckert, 2004), and in the visual system, no reduction in structural connectivity was observed in the LGN-V1 pathway (Müller-Axt et al., 2017).

We also performed the following control and exploratory analyses: (i) We analyzed the left mPT-MGB connection via A1 by performing an ANOVA with the factors "group" (neurotypicals versus dyslexics) and "track" (left mPT-MGB versus left mPT-MGB-via-A1). This was to rule out that a putative left mPT-MGB connectivity reduction in dyslexics is caused by alterations in left A1-MGB pathway. (ii) We assessed group effects in the MGB-IC connection by performing two-sample t-tests as well as an ANOVA with the factors "hemisphere" (left versus right) and "group" (neurotypicals versus dyslexics). This was to address the finding of several studies, which associated dyslexia with alterations of auditory brainstem structures (Chandrasekaran et al., 2009; Chandrasekaran and Kraus, 2010).

We first performed all analyses with ROIs that contained both grey and white matter voxels. The extension of grey matter regions to the surrounding white matter clusters is common practice in diffusion tensor imaging based probabilistic tractography analyses (Park et al., 2004; Soares et al., 2013; Thomas et al., 2014). Since grey matter and cerebro-spinal fluid have low anisotropy, tracking only from and to grey matter voxels (instead of white matter voxels adjacent to the grey matter region of interest), yields less robust probabilistic tractography results (Jones et al., 2013). Importantly, no group effect in the size of MGB volumes was observed for ROIs that contained both grey and white matter voxels (see Table 2). Group effects in the volumes of seed masks (i.e., mPT, see Table 3) were corrected within the tractography analysis pipeline, by accounting for the number of voxels in the respective seed mask (i.e., the Connectivity Index was calculated in a way that the number of streamlines from a given seed that reached the target was corrected by the number of voxels in the respective seed mask, see section 7). In addition, we performed tractography analyses with the white matter ROIs (cf. Anwander et al., 2007; Blank et al., 2011) for which no group differences were found in left-hemispheric subcortical ROIs (Table 2) and cerebral cortex ROIs (Table 3). We did not further analyze tractography results from grey matter ROIs (Behrens and Johansen-Berg, 2005), because the estimated tracks could not be considered reliable (i.e. there were not more than 10 streamlines reaching the target ROI): in 2 neurotypicals and 8 dyslexics did not show reliable estimates for the connection left mPT-MGB, and 4 dyslexics did not show reliable estimates for the connection left A1-MGB.

Effect sizes for the analyses were calculated using eta squared ($\eta^2$) (Cohen, 1973) for ANOVAs, and Cohen's $d_s$ (Cohen, 1988) for two-sample t-tests.

### 8.2 Correlation of connectivity measures with behavioral scores

To test our second hypothesis, i.e. that lower reading skills in dyslexics are associated with weaker left mPT-MGB connectivity strength, we correlated the respective connectivity measure with two behavioral scores: the RANln and reading comprehension. The RAN is a measure of reading fluency (for review, see Norton and Wolf, 2012), which previously correlated with atypical fMRI responses in the left MGB of dyslexics (Díaz et al., 2012), as well as cortico-subcortical structural connectivity strength in the visual pathway (Müller-Axt et al., 2017). Lower reading comprehension scores in dyslexics were also associated with atypical left MGB BOLD signal changes (Díaz et al., 2012). We hypothesized that weaker left mPT-MGB connectivity is associated with lower RANln and reading comprehension scores, and Bonferroni-corrected for the four conducted t-tests on the Pearson's correlations: i.e. the correlations between the two behavioral scores of interest, and the connectivity indices from the two analyses with (i) ROIs containing grey and white matter voxels, as well as (ii) ROIs containing white matter voxels only. We report 95% confidence intervals from bootstrapping with 1000 sampling rounds together with parametric statistical tests.

## Results

### 1. Left mPT-MGB connectivity is reduced in dyslexics in comparison to neurotypicals

The connection between the left mPT and the left MGB could be reliably estimated in all (N=24) participants for ROIs including white and grey matter, with at least 10 of the generated sample streamlines reaching the target mask. One neurotypical participant was excluded due to a lower-bound outlier in the left mPT-MGB connection.

To test our hypothesis that dyslexia is associated with reduced connectivity between left mPT and left MGB, we analyzed the differences in left mPT-MGB connectivity indices between dyslexic and neurotypical participants by means of a two-sample t-test. In accordance with our hypothesis, there were higher connectivity indices for neurotypicals than dyslexics in the left mPT-MGB connection (t(21)=3.27, p=.003, $d_s$=1.378) (Figure 4). To control for potential biases due to group differences in the mPT volumes (Table 3), this result was validated in analyses using white matter ROIs (see section 4).

[Please insert Figure 4]

### 2. Is the white matter reduction between mPT and MGB specific to the left hemisphere?

We next tested whether dyslexics' reduction in mPT-MGB connectivity was specific to the left hemisphere. We assumed a specificity in connectivity reduction to the left hemisphere based on the findings that the left MGB showed atypical fMRI responses in dyslexics (Díaz et al., 2012), and that post-mortem studies on brains of dyslexics found alterations in the left but not in the right MGB (Galaburda et al., 1994; cf. Stein, 2001). Also in the visual modality, the left-hemispheric cortico-thalamic connectivity was reduced in dyslexics as compared to neurotypicals, while no equivalent effect emerged in the right hemisphere (Müller-Axt et al., 2017). We calculated a 2x2 mixed-effects ANOVA with the factors "hemisphere" (left versus right) and "group" (neurotypicals versus dyslexics) for ROIs containing both grey and white matter. Again, to control for potential biases due to volume size, these results were validated using ROIs that contained white matter volume proportions only (see section 4). The connection between right mPT and right MGB could be reliably estimated in all 24 participants. The ANOVA showed a significant hemisphere x group interaction ($F(1,21)=12.26$, $p=.002$, $\eta^2=.369$) (Figure 5). Post-hoc t-tests revealed no significant difference in the right-hemispheric mPT-MGB connection between dyslexics and neurotypicals ($t(22)=.647$, $p=.524$, $d_s=.264$) (cf. Figure 5), indicating that the interaction was driven by the significant group effect (dyslexics versus neurotypicals) in the connection between the left mPT and the left MGB ($t(21)=3.27$, $p=.003$, $d_s=1.378$).

[Please insert Figure 5]

3. **Is dyslexics' reduction in white matter connectivity specific to the left mPT-MGB connection, but not present in the A1-MGB connection?**

When comparing the left mPT-MGB connection with the left A1-MGB connection, we expected no reduction in connectivity strength in the A1-MGB connection for two reasons: first, human neuroimaging as well as postmortem brain analyses did not show any evidence of dyslexia related alterations in A1 (for review, see Eckert, 2004), and second, there was no white matter connectivity reduction for dyslexics in the visual modality between V1 and the LGN (Müller-Axt et al., 2017). These two findings suggested that dyslexia related structural alterations might not be reflected in cortico-thalamic connectivity involving primary sensory cortices.

The connection between A1 and MGB could be reliably estimated in all 24 participants in both hemispheres (Figure 6). There was a marginally significant interaction in the 2x2 mixed-effects ANOVA with the factors "group" (neurotypicals versus dyslexics), and "seed region" (left A1 versus left mPT) with a medium effect size (F(1,21)=3.71, p=.068, $\eta^2$=.150). In contrast to the left mPT-MGB connection (Figure 4, Figure 5), the connectivity strength of the left A1-MGB pathway was not significantly different between groups (t(22)=1.49, p=.150, $d_s$=.608). There was also no hemisphere x group interaction in the A1-MGB connection (F(1,22)=1.08, p=.310, $\eta^2$=.047) (Figure 6).

[Please insert Figure 6]

### 4. Replication with ROIs that include white matter only

In addition to our analyses that included ROIs consisting of both grey and white matter, we replicated our hypothesis-driven analysis outcomes with ROIs that only contained white matter voxels. This was a) to control for potential biases due to differences in grey and white matter volume proportions across ROIs, b) to control for potential biases due to differences in overall ROI volume. There were no group effects in the size of left MGB volumes for ROIs that contained both grey and white matter (Table 2); however, we did observe a group difference in the size of the left mPT (Table 3). It is unlikely that this volume difference caused the group effect in left mPT-MGB connectivity strength, since streamline counts in connectivity analyses were corrected for the size of the seed region, i.e., the left mPT volume. Nevertheless, we aimed for replication of the group effect in left mPT-MGB connectivity strength with ROIs that did not show volume differences between groups.

Analyses with white matter ROIs revealed reliable tracking results in all participants (N=24), and no group differences in the sizes of the left MGB volume (Table 2), or in the size of the left mPT mask (Table 3). In accordance with our first hypothesis, we again found weaker left mPT-MGB connectivity strength for dyslexics than neurotypicals (t(21)=2.51, p=.019, $d_s$=1.03).

Also the further tests on the specificity of the mPT-MGB group effect showed qualitatively similar results as reported for the combined white and grey matter ROIs. There was a significant hemisphere x group interaction (F(1,22)=7.10, p=.014, $\eta^2$=.244), driven by the reported group effect in left mPT-MGB connection, which was absent in the right hemisphere (p=.982). This was the case although the white matter ROI of the right MGB was smaller in dyslexics than in neurotypicals. The results are in line with analyses on masks that contained both grey and white matter voxels (see Figures 4 and 5). White-matter ROIs also did not reveal any significant group effect in the A1-MGB connection, and the interaction from an 2x2 mixed-effects ANOVA with the factors "group" (neurotypicals versus dyslexics), and "seed region" (left A1 versus left mPT) was close to significance (F(1,22)=4.24, p=.051, $\eta^2$=.162), i.e., similar to results for the combined white and grey matter ROIs (Figure 6).

## 5. Control and exploratory analyses

### 5.1 Analysis of the left mPT-MGB connection via A1

We tested whether dyslexia related alterations in left mPT-MGB connectivity may originate from dysfunctional tracks that go via A1. While there is evidence for direct cortico-thalamic connectivity between higher-level auditory cortices, such as the mPT, and the MGB (Winer et al., 2001), it has been also suggested that some projections between the MGB and higher-level auditory cortices go via A1 (Rouiller et al., 1991; Lee, 2013). Thus, we compared the left mPT-MGB track via A1 with the direct connection between the left mPT and left MGB.

For tractography analyses via A1, the left mPT was set as seed region, the left A1 as waypoint, and the left MGB as second waypoint and termination mask. The tract via A1 showed overall weak connectivity strength, compared to the direct left mPT-MGB connection: two neurotypicals and one dyslexic were excluded due to absence of reliable connections. We ran a 2x2 mixed-effects ANOVA with the factors "group" (neurotypicals versus dyslexics) and "track" (left mPT-MGB versus left mPT-MGB-via-A1) to assess whether group effects primarily emerged in the direct connection between left mPT and left MGB. A significant track x group interaction (F(1,19)=6.28, p=.021, $\eta^2$=.249) showed dyslexics' reduction in connectivity strength in the direct left mPT-MGB connection (t(21)=3.27, p=.003, $d_s$=1.378; Figures 4, 5, 6), but not in the track via A1 (t(21)=.80, p=.427, $d_s$=.330).

**5.2 Analysis of MGB-IC connectivity**

Several studies (Chandrasekaran et al., 2009; Chandrasekaran and Kraus, 2010) have associated dyslexia with alterations of auditory brainstem structures. We therefore explored group differences in the connectivity between the IC and the MGB, and ran independent tracking analyses for both directions (MGB-to-IC and IC-to-MGB) within each hemisphere. Connections between the IC and the MGB showed reliable tracking results with at least 10 of the generated sample streamlines reaching the target mask for all participants (Figure 7A). We analyzed group differences in the averaged IC-to-MGB and MGB-to-IC tracking results in the left hemisphere by means of a two sample t-test, as well as hemispheric differences in an ANOVA with the factors "hemisphere" (left versus right) and "group" (neurotypicals versus dyslexics). However, no main effect or interaction with the factors group or hemisphere emerged (all p-values > .182) (Figure 7B). We obtained qualitatively similar results for the ROIs based on the alternative left IC coordinate (see methods).

[Please insert Figure 7]

**6. Correlation of left mPT-MGB connectivity strength with reading skills**

Our second hypothesis was that lower scores in the RANln, as well as in reading comprehension, are accompanied by weaker connectivity strength between the left mPT and the left MGB in dyslexics. We focused on this connection because it showed reliable group differences between dyslexics and neurotypicals across our analyses, in line with our hypothesis.

We Bonferroni corrected for the four conducted tests, i.e. the correlations between the two behavioral scores of interest, the RANln score and the reading comprehension score, and connectivity indices from analyses with ROIs containing grey and white matter voxels or white matter voxels only. Against our hypotheses, there was no significant correlation between left mPT-MGB connectivity strength and the RANln score in dyslexics (all p-values p>.05). Dyslexics also did not show a significant correlation between left mPT-MGB connectivity strength and the reading comprehension score (all p-values p>.05); there was, however, a negative correlation of the RANln score with mPT-MGB connectivity strength in neurotypical readers for analyses with ROIs that contained white matter voxels only (r=-.798, p=.002, 95% CI [-.310, -.979]) (Figure 8). This indicates that stronger connectivity strength between left mPT and left MGB was associated with higher levels of reading fluency, as measured by faster rapid naming skills, in neurotypical readers. Despite a similar trend, no Bonferroni-corrected significant correlation was observed for analyses with ROIs containing both grey and white matter voxels (r=-.553, p=.077).

[Please insert Figure 8]

## Discussion

In the current study, we found three key results. First, there was a reduction in connectivity strength between the left mPT and the left MGB in dyslexics compared to neurotypical readers, in line with our hypothesis. Second, dyslexics showed no reduction in structural connectivity in the right hemisphere, or in any other analyzed section of the auditory pathway. Third, our hypothesis that connectivity strength between the left mPT and the left MGB in dyslexics correlates with measures of reading fluency and reading comprehension was not confirmed. Instead, left mPT-MGB connectivity strength positively correlated with reading fluency in neurotypical readers. The results provide first evidence for the long-standing hypothesis (cf. Galaburda et al., 1994) that dyslexia related functional (Díaz et al., 2012) and structural (Galaburda et al., 1994; cf. Stein, 2001) alterations in subcortical regions of the auditory pathway (i.e., in the left MGB) are accompanied by reduced connectivity to specific higher-level auditory cortices.

The hypotheses for our study were based on the findings of two previous studies: an atypical task-dependent modulation of the left MGB was observed for dyslexics during a phonological task in contrast to a control task (Díaz et al., 2012), and dyslexics showed reduced connectivity between left V5/MT and the left LGN (Müller-Axt et al., 2017). The findings of the present study from the auditory modality mirror the previous findings in the visual modality (Müller Axt et al., 2017). In that study a reduction in the left-hemispheric pathway between the visual sensory thalamus (LGN) and the visual motion-sensitive cortex (MT/V5) was found in dyslexics as compared to neurotypicals. In analogy, we here found a specific reduction in the left-hemispheric pathway between the MGB and the mPT in dyslexics as compared to neurotypical readers. The results could not be explained by a general reduction in connectivity strength in dyslexia as there were no group differences in the right mPT-MGB pathway, the A1-MGB pathway, or the MGB-IC pathway. Left-right hemispheric asymmetries in the auditory modality have been found in other studies on the functional (Belin et al., 1998; Schönwiesner et al., 2007) as well as structural (Mišić et al., 2018) level. In dyslexics, a specificity of the mPT-MGB pathway reduction to the left hemisphere is in agreement with its specialization for speech processing, and the speech perception-related deficits found in the dyslexia population, such as phonological processing difficulties (for example, Snowling and Melby-Lervåg, 2016).

The here reported group difference in left mPT-MGB connectivity between dyslexics and neurotypicals is large ($d_s$=1.37), similarly to the effect size in the visual domain in the left V5/MT-to-LGN connection ($d_s$=1.28) (Müller-Axt et al., 2017). Ramus et al. (2017) attempted to estimate the effect size of neuroanatomical differences in developmental disorders based on a voxel-based morphometry study in dyslexics (Eckert et al., 2016), and a meta-analysis of cortical thickness in subjects with autism spectrum disorder (Haar et al., 2016), concluding that effect sizes above 0.5 in the field of dyslexia research might be inflated. However, whether or not the here reported effects, as observed in two sensory domains, are inflated due to a modest sample size (cf. Button et al., 2013) can only be answered by future (large sample) diffusion tensor imaging studies on developmental disorders.

Animal tracing studies detected independent fiber tracks between association auditory cortices and the MGB. Those studies discovered that the majority of neural fibers *directly* connect sub-nuclei of the MGB with different parts of the auditory association cortex (Hackett, 2011). Fiber tracing analyses in cats revealed more than 12 independent connections from parts of the primary auditory cortex, as well as from auditory association cortices, to the MGB (Bajo et al., 1995; Winer et al., 2001). The homology of auditory cortex regions in animals and humans is a matter of debate (cf. Zatorre and Belin, 2001) and it is therefore unclear which of the association auditory cortex regions that are directly connected to the MGB might correspond to the PT. However, since parts of the PT (Moerel et al., 2014; Saenz and Langers, 2014; Shrem and Deouell, 2014) are tonotopically organized in humans, and direct fibre projections in both directions between all tonotopically organised regions and the MGB have been reported in animals (Bajo et al., 1995; Winer et al., 2001; Lee and Winer, 2008) it is likely that that direct fiber connections exist between the MGB and the PT in humans.

Evidence from animal models and human neuroimaging suggests that both the PT and MGB are sensitive to spectotemporally complex sound. First, animal research on frequency modulated sweeps, which mimic the fast frequency modulations in speech, has shown that especially the ventral part of the MGB is sensitive to the direction of those sweeps, i.e. changes from low to high frequencies and vice versa (cf. Wenstrup, 1999; Lui and Mendelson, 2003). Second, a specific encoding of complex spectotemporal characteristics in the PT has been shown in the context of auditory motion detection paradigms (for review, see Griffiths and Warren, 2002; Alink et al., 2012). Auditory and visual motion processing have been considered key components for the acquisition of reading and writing skills (Witton et al., 1998; Talcott et al., 2002; cf. Joo et al., 2017). Both types of motion processing were found to be deficient in children and adults with dyslexia (Stein, 2001; Amitay et al., 2002; Facoetti et al., 2003).

Visual and auditory motion processing cortices, i.e. MT/V5 and mPT, might modulate their corresponding sensory thalamic nuclei during the processing of spoken speech. Evidence for task-dependent modulation of sensory thalamic nuclei has been provided by means of fMRI: task-dependent modulation of the left MGB was stronger in a speech sound recognition task, as compared to a voice identity recognition task on the same stimuli (von Kriegstein et al., 2008; for similar findings on the LGN, see Díaz et al., 2018). This suggested that the MGB preferentially engages in the processing of phonological information in comparison to slowly occurring auditory features, such as required for voice identity detection. Developmental dyslexia has been associated with deficient task-dependent modulation of the left MGB specifically during phonological processing (Díaz et al., 2012). Thus, our finding of reduced left mPT-MGB connectivity in dyslexics might reflect a reduction in cortico-thalamic feedback connections between mPT and MGB. But this is speculative at the current stage, since probabilistic tracking analyses do not allow inferences on the direction of resolved connections.

Our analyses suggested that the left mPT-MGB connectivity may at least partially contribute to reading performance. This was supported by a significant correlation between lower reading fluency measures, as operationalized by the RANln, and weaker left mPT-MGB connectivity strength in neurotypical readers. The RANln has been linked to both the reading abilities of dyslexics (Semrud-Clikeman et al., 2000), as well as neurotypical readers (Miller et al., 2006). A meta-analysis with a sample size of > 2.000 found correlations between RANln and reading skills to be similarly strong in poor and neurotypical readers (Swanson et al., 2003). In the present study correlation between RANln and left mPT-MGB connectivity strength was found only in neurotypical readers. Similar effects were previously reported for dyslexics: slower performance in the RANln was associated with weaker connection strength in the visual domain between left V5/MT and left LGN (Müller-Axt et al., 2017), and with reduced speech task-dependent modulation in the left MGB (Díaz et al., 2012). It is interesting that all studies reported an association of the respective neuroscience measures with RANln scores. However, it is unclear why two of the studies found associations in the dyslexia group and the present study only in the neurotypical group. Furthermore the results of the correlations in the present study have to be taken with caution as they did not correspond to our a priori hypothesis and were present only in one of the analyses (i.e., with white matter ROIs only).

We conclude that for a comprehensive understanding of dyslexia mechanisms, cortico-thalamic structural alterations need to be taken into account. Specifically, the current study provided evidence for such alterations in the auditory system, while equivalent findings have been previously reported for the visual domain (Müller-Axt et al., 2017). This implies that developmental dyslexia needs to be discussed within a multi-sensory framework that also includes sub-cortical brain structures involved in earliest visual and auditory processing.

## Acknowledgments

We are grateful to our participants for their time and effort to participate in this study. We thank Domenica Wilfling, Elisabeth Wladimirow, and Florian Hintz for providing the structural MRI data and the dyslexia diagnostic scores. We thank Alfred Anwander and Christa Müller-Axt for their helpful advice on our tracking analysis protocols. The work was supported by a Max

Planck Research Group grant and an ERC-Consolidator Grant (SENSOCOM, 647051) to K.v.K. The authors declare no competing financial interests.

# Figure Legends

**Figure 1.** The left side displays the statistical parametrical map of the localizer contrast "Sentences - Silence". The color bar represents t-values. The crosshair over the zoomed-in medial geniculate body responses indicates the statistic peak location from the localizer contrast "Sentences - Silence", used for definition of ROIs. The right side shows the ROIs centered at the statistic peak location of the localizer contrast "Sentences - Silence". The maps and ROIs are superimposed on the same section of the MNI152 structural T1 volume.

**Figure 2.** Localization of the medial geniculate body (MGB) masks on the individual subject T1 brain, presented for six randomly chosen dyslexic and neurotypical participants, respectively.

**Figure 3**. Regions of interest (ROIs) in MNI standard space, superimposed on sections of the MNI152 structural T1 volume.

**Figure 4**. **Panel A**: Averaged probabilistic white matter connectivity for neurotypicals and dyslexics between the left motion-sensitive planum temporale (mPT) and the left medial geniculate body (MGB) (green). The log-normalized and averaged tracks are presented in MNI standard space, and thresholded to the same minimum value of 0.08. **Panel B**: Mean connectivity strength of the left mPT-MGB connection for neurotypicals and dyslexics, respectively. Error bars indicate ± 1 SEM.

**Figure 5**. Mean connectivity strength for neurotypicals and dyslexics between the motion-sensitive planum temporale (mPT) and the medial geniculate body (MGB). Error bars indicate ± 1 SEM.

**Figure 6**. **Panel A**: Averaged probabilistic white matter connectivity for neurotypicals and dyslexics between the motion-sensitive planum temporale (mPT) and the medial geniculate

body (MGB) (dark green), and between the primary auditory cortex (A1) and the MGB (blue). The log-normalized and averaged tracks are presented in MNI standard space, and thresholded to the same minimum value of 0.08. **Panel B**: Mean connectivity strength of the mPT-MGB and the A1-MGB connectivity for neurotypicals and dyslexics, respectively. Error bars indicate ± 1 SEM.

**Figure 7**. Averaged probabilistic white matter connectivity for neurotypicals and dyslexics between the medial geniculate body (MGB) and the inferior colliculus (IC). **Panel A**: The log-normalized and averaged tracks are presented in MNI standard space, and thresholded to the same minimum value of 0.50. **Panel B**: Mean connectivity strength of tracks between IC and MGB for neurotypicals and dyslexics, respectively. Error bars indicate ± 1 SEM.

**Figure 8**. Correlation between the rapid automatized naming score for letters and numbers (RANln) and the connectivity index of the white matter pathway between the left motion-sensitive planum temporale (mPT) and the left medial geniculate body (MGB). A significantly negative correlation emerged for neurotypical participants, suggesting that stronger mPT-MGB connectivity was associated with faster rapid automatized naming of letters and numbers.

## Table Legends

**Table 1**. Social demographic and cognitive measures (mean ± SD) of 12 male dyslexics, and 12 male neurotypical participants. The raven matrices test (mean = 100, SD = 15), the spelling test (mean = 100, SD = 10), as well as the reading speed and comprehension tests (mean = 50, SD = 10) are all based on standard scores.

**Table 2**. The size of subcortical regions of interest (ROI) in diffusion space. Spherical masks were defined around the statistical maxima of the functional localizer in the anatomical locations of the left and right Medial Geniculate Body (MGB) (left [-15, -28, -5], and right [12, -28, -8] in MNI space), as well as the left and right Inferior Colliculus (IC) (left [-6, -34, -8], and right [6, -37, -5] in MNI space), respectively.

**Table 3**. The size of cortical regions of interest (ROIs) in diffusion space. Peak coordinates within the primary auditory cortex (A1) (left [-51, -16, 4], and right [42, -22, 7] in MNI space) and planum temporale (PT) (left [-53, -31, 12], and right [54, -29, 14] in MNI space) were intersected with thresholded atlas masks from the Juelich Histological Atlas (Eickhoff et al., 2005) and the Harvard-Oxford-Atlas (Desikan et al., 2006), respectively.

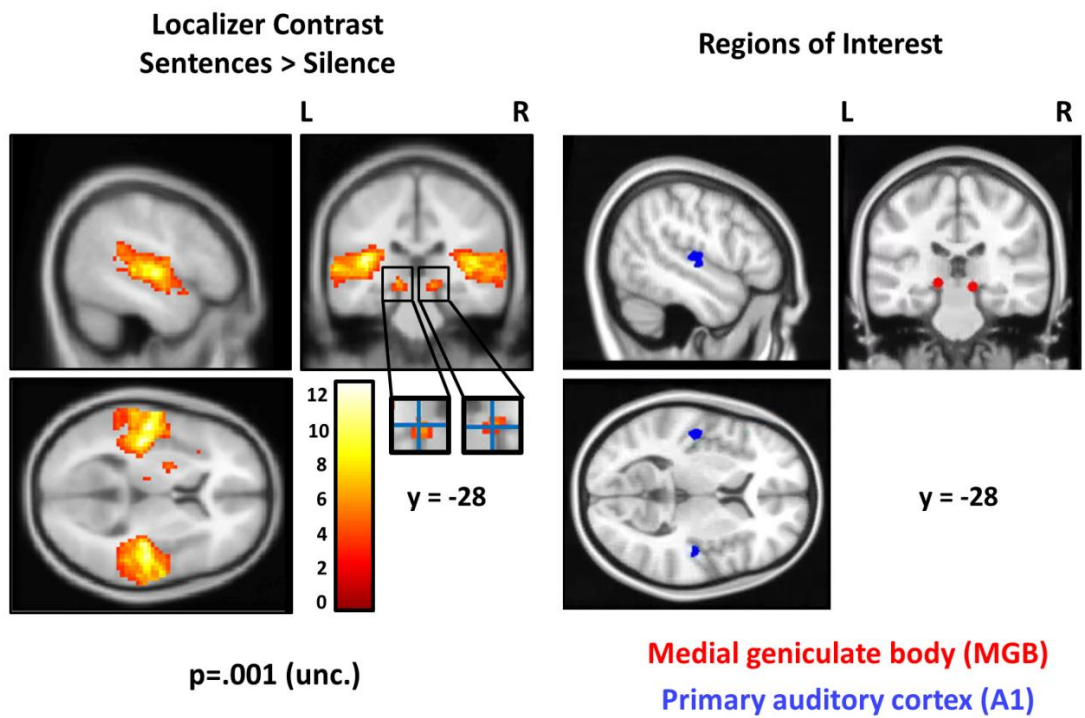

**Localizer Contrast Sentences > Silence**

**Regions of Interest**

y = -28

p=.001 (unc.)

**Medial geniculate body (MGB)**
**Primary auditory cortex (A1)**

**Figure 1.** The left side displays the statistical parametrical map of the localizer contrast "Sentences - Silence". The color bar represents t-values. The crosshair over the zoomed-in medial geniculate body responses indicates the statistic peak location from the localizer contrast "Sentences - Silence", used for definition of ROIs. The right side shows the ROIs centered at the statistic peak location of the localizer contrast "Sentences - Silence". The maps and ROIs are superimposed on the same section of the MNI152 structural T1 volume.

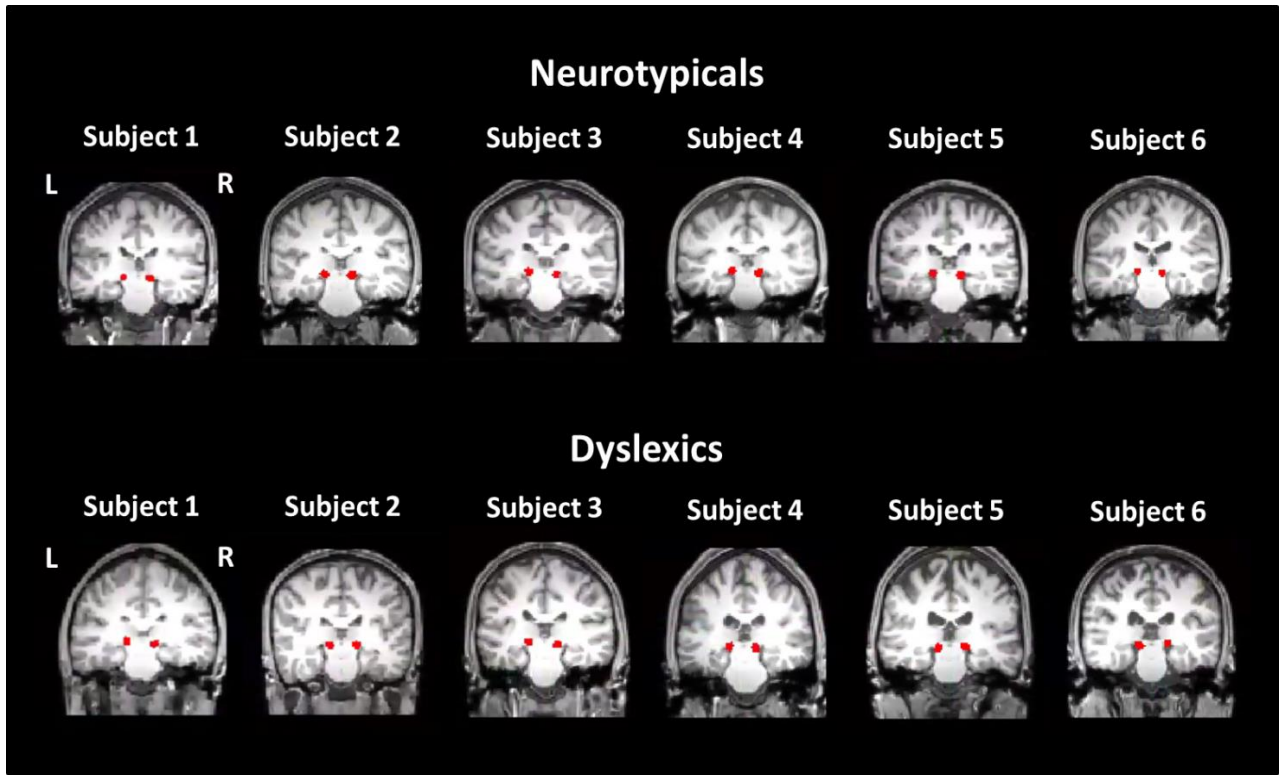

**Figure 2.** Localization of the medial geniculate body (MGB) masks on the individual subject T1 brain, presented for six randomly chosen dyslexic and neurotypical participants, respectively.

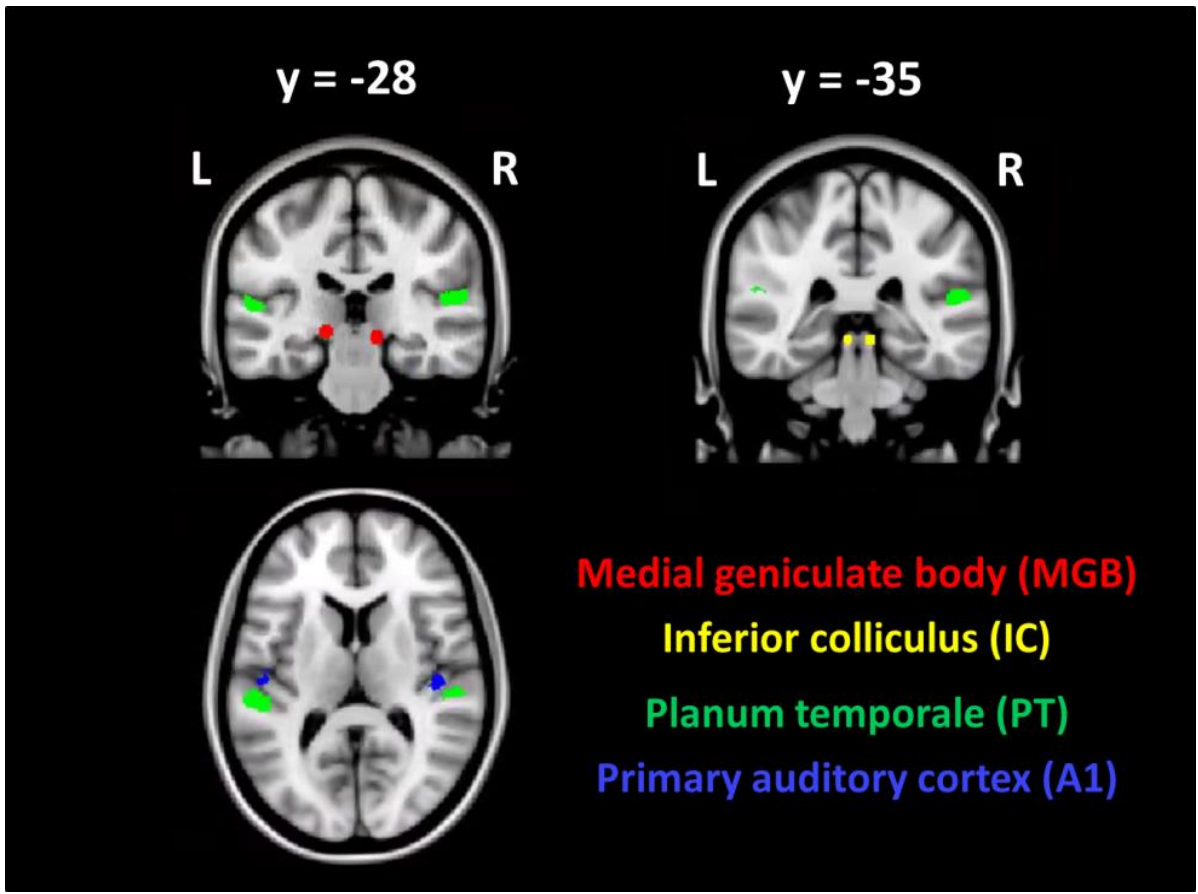

**Figure 3**. Regions of interest (ROIs) in MNI standard space, superimposed on sections of the MNI152 structural T1 volume.

# Left mPT - MGB

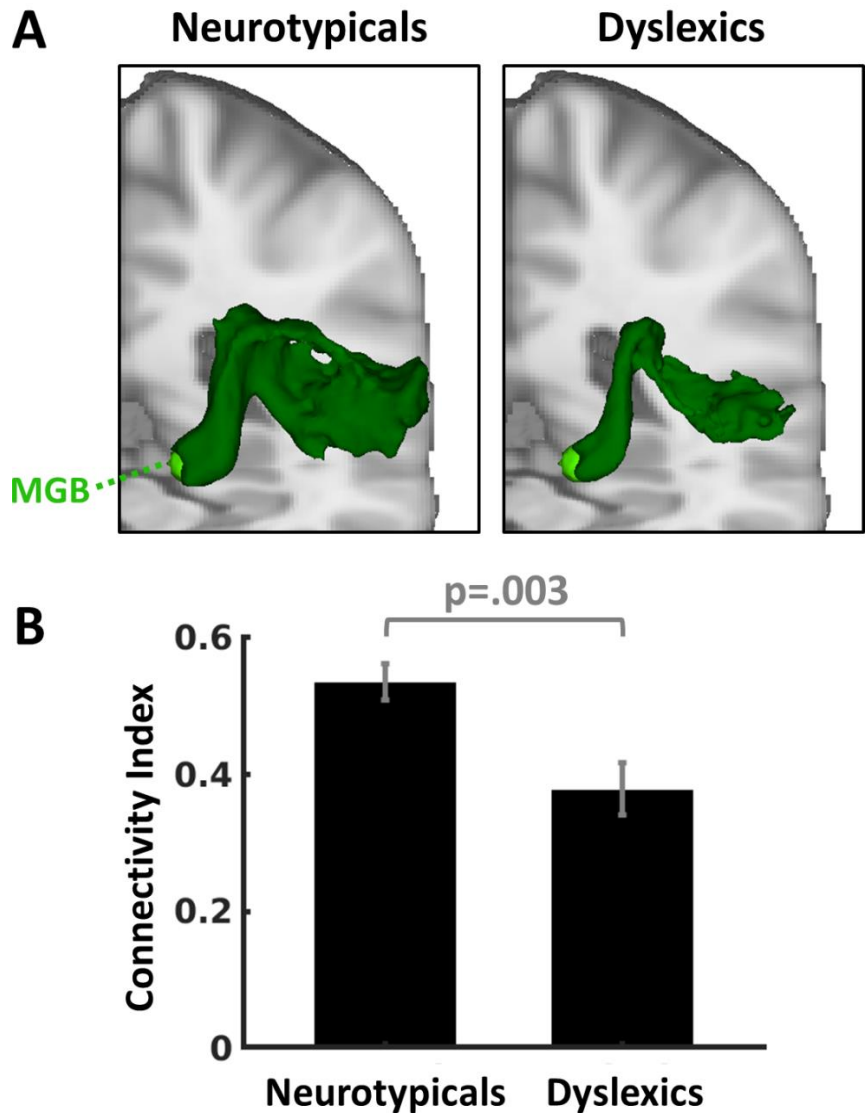

**Figure 4**. **Panel A**: Averaged probabilistic white matter connectivity for neurotypicals and dyslexics between the left motion-sensitive planum temporale (mPT) and the left medial geniculate body (MGB) (green). The log-normalized and averaged tracks are presented in MNI standard space, and thresholded to the same minimum value of 0.08. **Panel B**: Mean connectivity strength of the left mPT-MGB connection for neurotypicals and dyslexics, respectively. Error bars indicate ± 1 SEM.

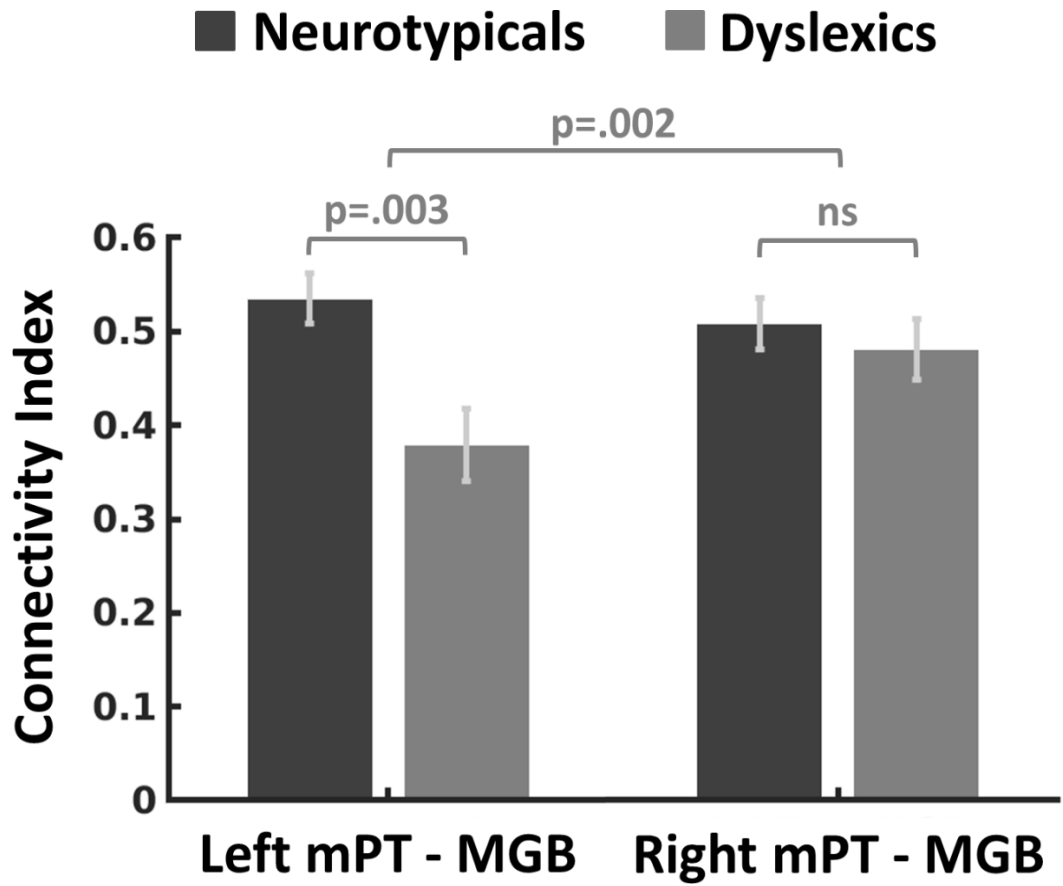

**Figure 5**. Mean connectivity strength for neurotypicals and dyslexics between the motion-sensitive planum temporale (mPT) and the medial geniculate body (MGB). Error bars indicate ± 1 SEM.

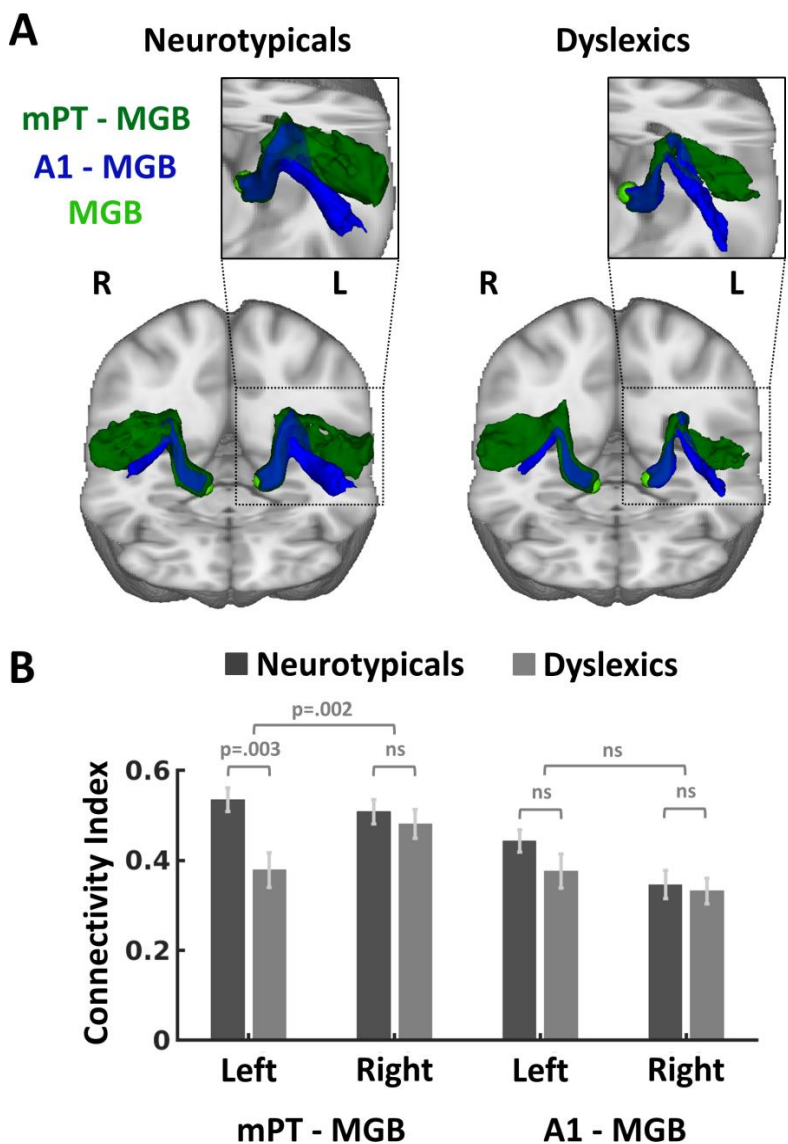

**Figure 6**. **Panel A**: Averaged probabilistic white matter connectivity for neurotypicals and dyslexics between the motion-sensitive planum temporale (mPT) and the medial geniculate body (MGB) (dark green), and between the primary auditory cortex (A1) and the MGB (blue). The log-normalized and averaged tracks are presented in MNI standard space, and thresholded to the same minimum value of 0.08. **Panel B**: Mean connectivity strength of the mPT-MGB and the A1-MGB connectivity for neurotypicals and dyslexics, respectively. Error bars indicate ± 1 SEM.

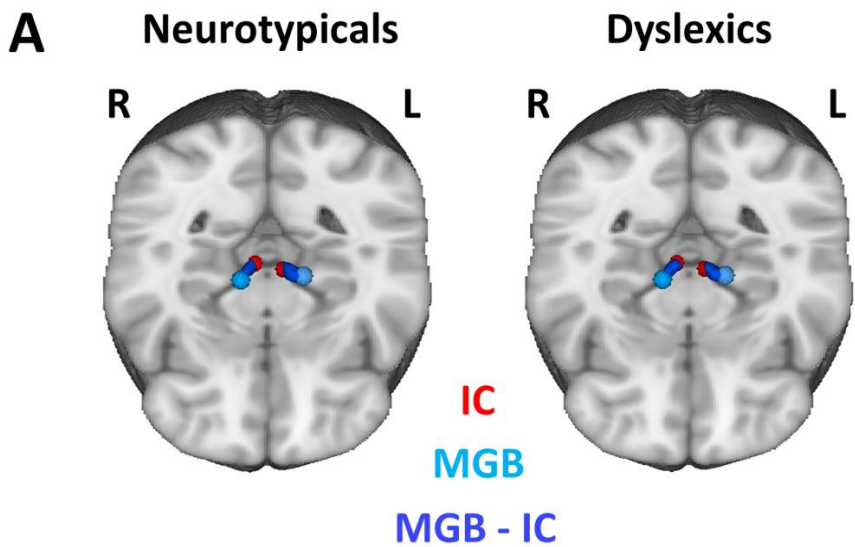

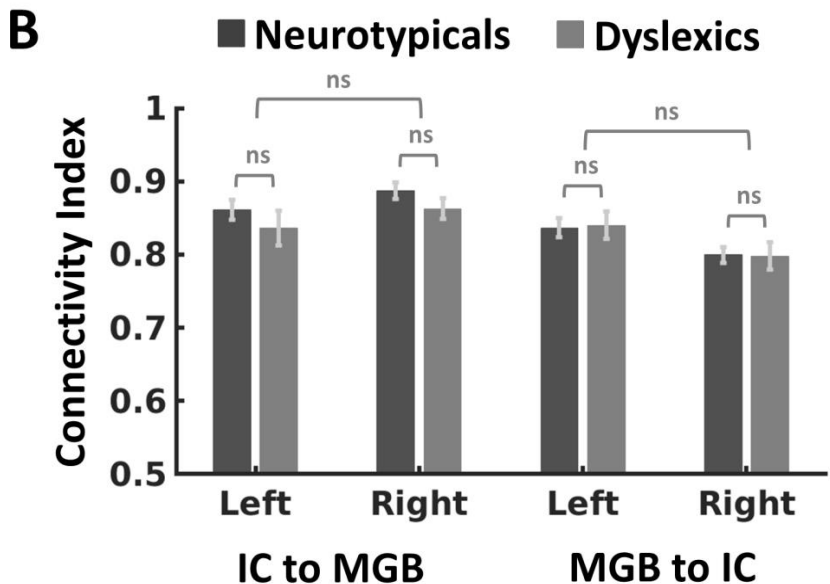

**Figure 7**. Averaged probabilistic white matter connectivity for neurotypicals and dyslexics between the medial geniculate body (MGB) and the inferior colliculus (IC). **Panel A**: The log-normalized and averaged tracks are presented in MNI standard space, and thresholded to the same minimum value of 0.50. **Panel B**: Mean connectivity strength of tracks between IC and MGB for neurotypicals and dyslexics, respectively. Error bars indicate ± 1 SEM.

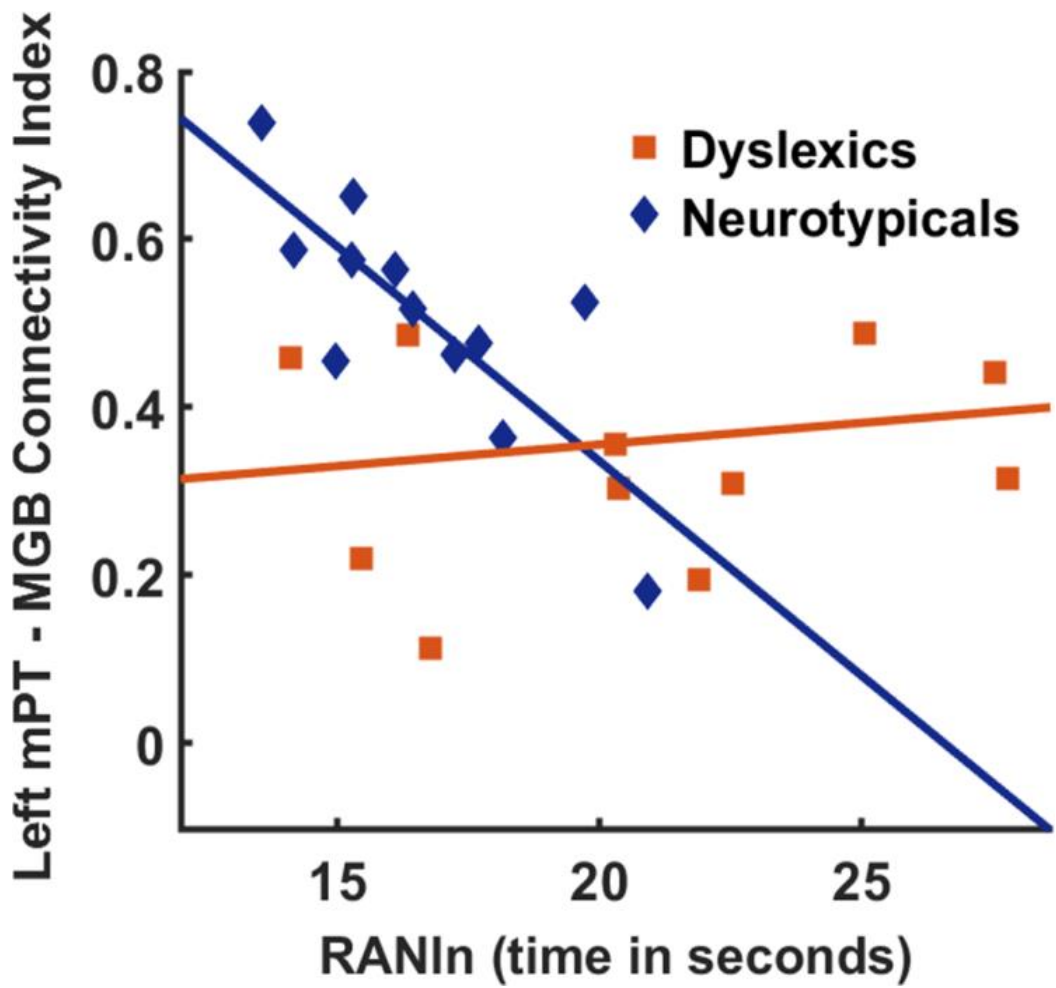

**Figure 8**. Correlation between the rapid automatized naming score for letters and numbers (RANln) and the connectivity index of the white matter pathway between the left motion-sensitive planum temporale (mPT) and the left medial geniculate body (MGB). A significantly negative correlation emerged for neurotypical participants, suggesting that stronger mPT-MGB connectivity was associated with faster rapid automatized naming of letters and numbers.

| | Neurotypical group (N = 12) | Dyslexia group (N = 12) | Two-sample t-tests neurotypicals - dyslexics |
|---|---|---|---|
| **Demographic data** | | | |
| Age in years | 23.7 ± 2.6 | 24.2 ± 2.3 | not significant (NS) |
| Handedness | right=11, left=1 | right=10, left= 2 | --- |
| Education | 11 undergrad. students, 1 high school diploma | 12 undergrad. students | --- |
| **Diagnostic tests** | | | |
| non-verbal IQ (Raven matrices) | 110.8 ± 12.8 | 101 ± 13.6 | NS |
| Spelling | 102.8 ± 5.6 | 83.1 ± 7.6 | t(22)=7.2, p<.001 |
| Reading speed | 58.3 ± 9.1 | 42.6 ± 6.5 | t(22)=4.9, p<.001 |
| Reading comprehension | 62.9 ± 7.7 | 47.4 ± 4.2 | t(22)=6.1, p<.001 |
| RAN numbers time (ms) | 16.8 ± 2.4 | 21.2 ± 6.1 | t(22)=2.3, p<.05 |
| RAN numbers errors (%) | 0.8 ± 1.3 | 0.2 ± 0.6 | NS |
| RAN letters time (ms) | 16.4 ± 2.6 | 20.3 ± 3.5 | t(22)=3.1, p<.01 |
| RAN letters errors (%) | 0.3 ± 1.2 | 0.3 ± 0.8 | NS |

**Table 1**. Social demographic and cognitive measures (mean ± SD) of 12 male dyslexics, and 12 male neurotypical participants. The raven matrices test (mean = 100, SD = 15), the spelling test (mean = 100, SD = 10), as well as the reading speed and comprehension tests (mean = 50, SD = 10) are all based on standard scores.

| Subcortical regions of interest | | | |
|---|---|---|---|
| ROI volumes in mm$^3$, mean ± SD | Neurotypical group (N = 12) | Dyslexia group (N = 12) | Two-sample t-tests neurotypicals - dyslexics |
| **Left Medial Geniculate Body (MGB)** | | | |
| Grey and white matter | 347.86 ± 32.30 | 364.20 ± 33.32 | t(22) = -1.22, p = .235 |
| Grey matter | 70.82 ± 34.07 | 117.34 ± 30.50 | t(22) = -3.52, p = .001* |
| White matter | 277.02 ± 59.87 | 246.85 ± 44.64 | t(22) = 1.39, p = .175 |
| **Right MGB** | | | |
| Grey and white matter | 349.53 ± 53.97 | 348.69 ± 24.29 | t(22) = .049, p = .961 |
| Grey matter | 115.25 ± 47.44 | 151.71 ± 35.80 | t(22) = -2.12, p = .045* |
| White matter | 234.28 ± 52.76 | 196.98 ± 27.86 | t(22) = 2.16, p = .041* |
| **Left Inferior Colliculus (IC)** | | | |
| Grey and white matter | 153.39 ± 26.57 | 172.25 ± 25.56 | t(22) = -1.77, p = .090 |
| Grey matter | 85.50 ± 28.21 | 103.92 ± 22.52 | t(22) = -1.76, p = .090 |
| White matter | 67.89 ± 19.48 | 68.31 ± 14.46 | t(22) = -.05, p = .952 |
| **Right IC** | | | |
| Grey and white matter | 138.72 ± 31.11 | 163.45 ± 26.39 | t(22) = -2.10, p = .047* |
| Grey matter | 119.85 ± 27.45 | 138.30 ± 35.00 | t(22) = -1.43, p = .165 |
| White matter | 18.86 ± 11.95 | 25.15 ± 14.05 | t(22) = -1.18, p = .250 |

**Table 2**. The size of subcortical regions of interest (ROI) in diffusion space. Spherical masks were defined around the statistical maxima of the functional localizer in the anatomical locations of the left and right Medial Geniculate Body (MGB) (left [-15, -28, -5], and right [12, -28, -8] in MNI space), as well as the left and right Inferior Colliculus (IC) (left [-6, -34, -8], and right [6, -37, -5] in MNI space), respectively.

| Cortical regions of interest | | | |
|---|---|---|---|
| ROI volumes in mm$^3$, mean ± SD | Neurotypical group (N = 12) | Dyslexia group (N = 12) | Two-sample t-tests neurotypicals - dyslexics |
| **Left motion-sensitive planum temporale (mPT)** | | | |
| **Grey and white matter** | 1369.22 ± 257.36 | 1147.93 ± 201.34 | t(22) = 2.34, p = .028* |
| **Grey matter** | 959.33 ± 235.43 | 804.68 ± 126.19 | t(22) = 2.00, p = .057 |
| **White matter** | 282.47 ± 95.22 | 232.60 ± 125.52 | t(22) = 1.09, p = .284 |
| **Right mPT** | | | |
| **Grey and white matter** | 1246.42 ± 158.86 | 1167.21 ± 185.40 | t(22) = 1.12, p = .273 |
| **Grey matter** | 856.23 ± 135.27 | 722.95 ± 144.25 | t(22) = 2.33, p = .029* |
| **White matter** | 194.04 ± 104.93 | 252.30 ± 87.55 | t(22) = -1.47, p = .153 |
| **Left primary auditory cortex (A1)** | | | |
| **Grey and white matter** | 760.67 ± 103.36 | 700.32 ± 128.31 | t(22) = 1.26, p = .217 |
| **Grey matter** | 510.89 ± 93.01 | 464.37 ± 85.16 | t(22) = 1.27, p = .214 |
| **White matter** | 249.78 ± 111.84 | 235.95 ± 66.14 | t(22) = .368, p = .715 |
| **Right A1** | | | |
| **Grey and white matter** | 594.29 ± 136.80 | 527.23 ± 111.20 | t(22) = 1.31, p = .201 |
| **Grey matter** | 453.05 ± 99.00 | 375.93 ± 96.52 | t(22) = 1.93, p = .066 |
| **White matter** | 141.23 ± 68.77 | 151.29 ± 50.17 | t(22) = -.409, p = .686 |

**Table 3**. The size of cortical regions of interest (ROIs) in diffusion space. Peak coordinates within the primary auditory cortex (A1) (left [-51, -16, 4], and right [42, -22, 7] in MNI space) and planum temporale (PT) (left [-53, -31, 12], and right [54, -29, 14] in MNI space) were intersected with thresholded atlas masks from the Juelich Histological Atlas (Eickhoff et al., 2005) and the Harvard-Oxford-Atlas (Desikan et al., 2006), respectively.